\documentclass[aps,twocolumn,floatfix]{revtex4}
\usepackage{amsmath,amsfonts,amssymb}
\usepackage{graphicx}
\usepackage{relsize}
\usepackage{color}
\usepackage[normalem]{ulem}
\def\0#1{\xout{#1}}
\def\1#1{{\color{red}#1}}
\def\2#1{{\color{blue}#1}}
\def\3#1{{\color{magenta}#1}}
\def\4#1{{\color{yellow}#1}}
\begin{document}

\title{Random walks across the sea: the origin of rogue waves?}

\author{Simon Birkholz}
\affiliation{Max-Born-Institut, Max-Born-Stra\ss e 2A, 12489 Berlin, Germany}

\author{Carsten Br\'ee}
\affiliation{Weierstra\ss-Institut, Mohrenstr.~39, 10117 Berlin, Germany}

\author{Ivan Veseli\'c}
\affiliation{Fakult\"at f\"ur Mathematik, Technische Universit\"at Chemnitz, Reichenhainer Str.~41, 09126 Chemnitz, Germany}

\author{Ayhan Demircan}
\affiliation{Leibniz-Universit\"at Hannover, Welfengarten 1, 30167 Hannover, Germany}

\author{G\"unter Steinmeyer}
\affiliation{Max-Born-Institut,  Max-Born-Stra\ss e\,2A, 12489 Berlin, Germany}
\date{\today}

\begin{abstract}
Ocean rogue waves are large and suddenly appearing surface gravity waves \cite{Windwave}, which may cause severe damage to ships and other maritime structures \cite{review1,review2}. Despite  years of research, the exact origin of rogue waves is still disputed \cite{linearvsnonlinear}. Linear interference of waves with random phase has often been cited as one possible explanation \cite{LonguetHiggins,Forristall},
but apparently does not satisfactorily explain the probability of extreme events in the ocean \cite{Stansell,Christou}.
Other explanations therefore suggested a decisive role of a nonlinearity in the system \cite{review1,review2,nonlinear,nonlinear2,Jannsen}. Here we show that linear interference \cite{Arecchi,linearmicrowave} of a finite and variable number of waves \cite{IEEE1,IEEE2} may very well explain the heavy tail in the wave height distribution. Our model can explain all prototypical ocean rogue waves reported so far, including the ``three sisters'' \cite{threesisters} as well as rogue holes \cite{roguehole}. We further suggest nonlinear time series analysis \cite{Grassberger,dimension} for estimation of the characteristic number of interfering waves for a given sea state. If ocean dynamics is ruled by interference of less than ten waves, rogue waves cannot appear as a matter of principle. In contrast, for larger numbers, their appearance is much more likely than predicted by parameterless models \cite{Forristall} or longterm observation \cite{Christou}. The pronounced threshold behavior of our model enables effective forecasting of extreme ocean waves.
\end{abstract}

\maketitle

\begin{figure*}[tb]
\centering
\includegraphics[width=12cm]{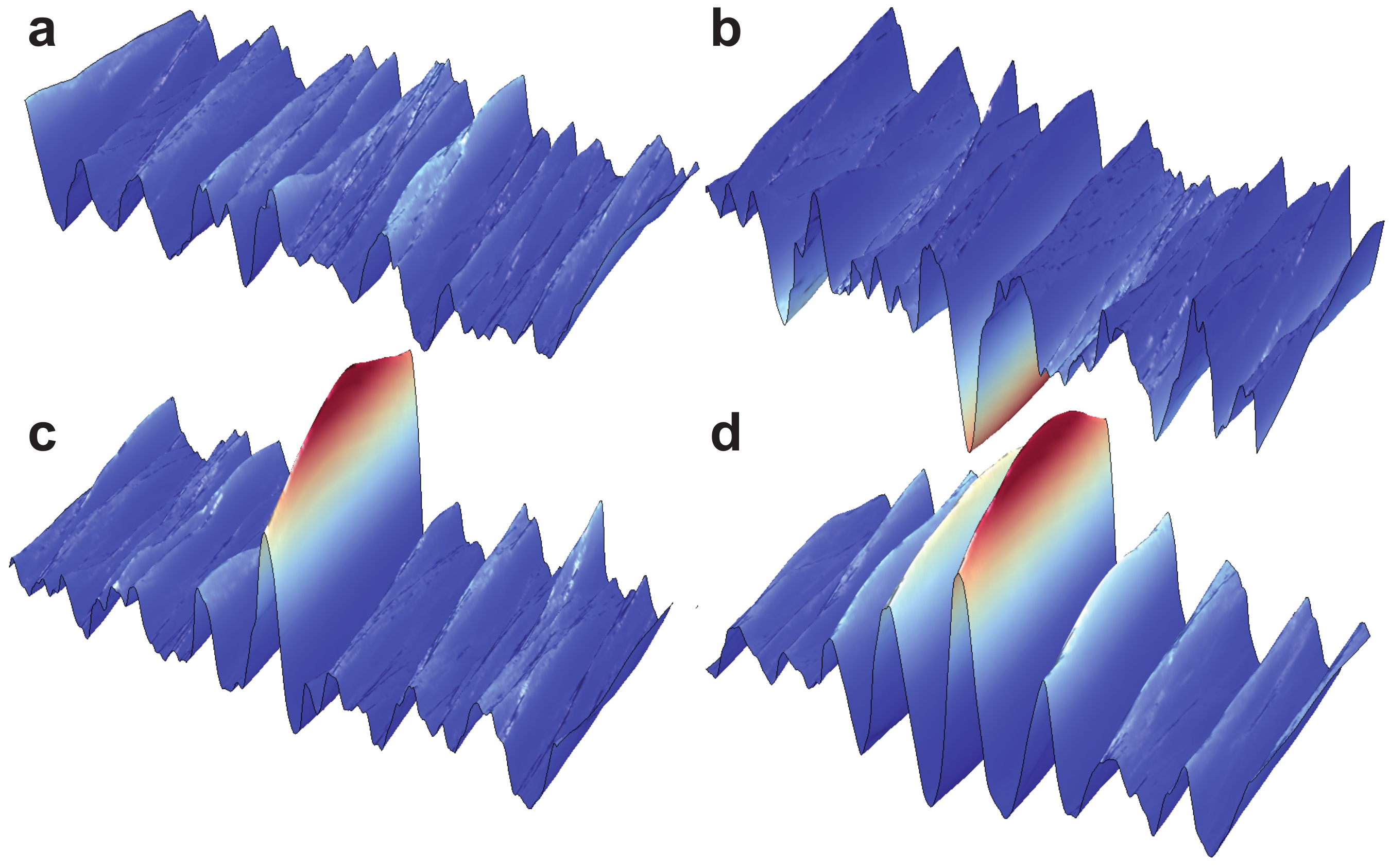}
\vspace{-1em}
\footnotesize{\caption{Simulation of the interference on a surface with a normally distributed angular spread of $N$ waves amounting to 50\,mrad. In this simulations, a nonlinear correction was included to reduce the depth of the troughs and steepen the wave crests. (a) Assuming random phases between the waves. A highly random wave pattern emerges with no wave significantly exceeding the significant wave height. $N=20$. (b) Simulation of a rogue hole, with synchronized phases adding up to negative interference at the central point of the propagation axis. $N=40$. (c) Same with phase flipped by $\pi$ and $N=20$, resulting in a single wave with extreme height $>2.5 H_{\rm S}$ Depending on the diffusion coefficient, leading and trailing troughs appear, similar to the observation of the Draupner event \cite{Draupner}. (d) Same with weaker phase diffusion. A rogue wave with structure reminiscent of the ``three sisters'' \cite{threesisters} emerges. \label{fig:1}}}
\vspace{-1em}
\end{figure*}

\noindent Surface gravity waves are a result of wind blowing over the ocean surface \cite{Windwave}. As wind directions differ across the ocean and in time, waves with different origin may start to interfere with each other at a given fixed location. In narrow confined waters like small ponds, one can sometimes observe the build-up of perfectly plane wavefronts across the entire surface width. In such a case, virtually no interference appears. In the much larger ocean system, in contrast, one anticipates an inextricable interference of many waves, which is typically treated in the limiting case of an infinite number of interfering waves \cite{LonguetHiggins,Forristall}. Mathematically modeling this interference process indicates a heavytail distribution of wave heights. Considering only linear interference, a Rayleigh distribution results \cite{LonguetHiggins}. Accounting for nonlinearities of the ocean system, the Tayfun distribution with an even more pronounced heavy tail emerges \cite{Tayfun}. Unfortunately, both distributions tend to significantly overestimate the appearance frequency of rogue events in the ocean \cite{Stansell,Christou}. In the following we will show that the heavy tail of the distribution can be adjusted by assumption of a finite number $N$ of interfering waves \cite{IEEE1,IEEE2}, as illustrated in Fig.~\ref{fig:1}. In particular, if $N$ drops below a certain threshold value, rogue waves become completely impossible whereas their likelihood rapidly increases with growing $N$ above the threshold. Moreover, the number $N$ is shown to be variable, depending strongly on location and weather conditions. Consequently, estimation of $N$ enables early warning of situations that are prone to rogue wave formation.

\begin{figure}[tb]
\centering
\includegraphics[width=7cm]{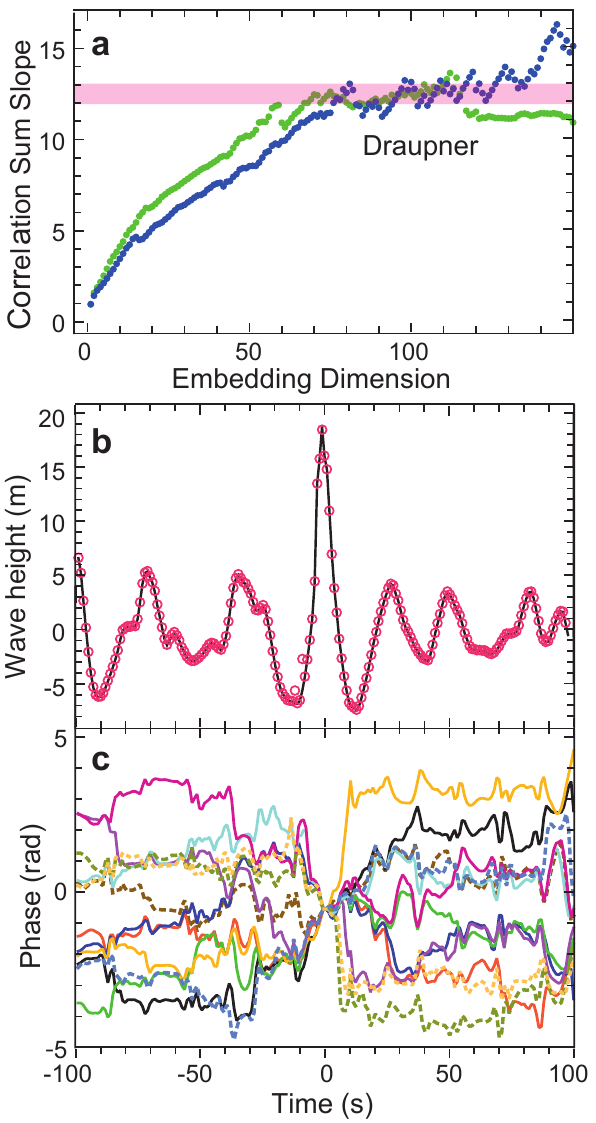}
\vspace{-1em}
\footnotesize{\caption{(a) Dimensional analysis of the Draupner wave record with the Grassberger-Procaccia method (see Methods section). Shown is the correlation sum slope as a function of embedding dimension for the original Draupner record (green) and a second measurement from the same campaign (blue). Both records contain a rogue event and exhibit a plateau of the slope for embedding dimensions ranging from 70 to 130. This results in an estimate for the information dimension $D$ of the record of 12 to 13, see pink shaded area. (b) Reconstruction of the Draupner event using the model assumption of an interference of 12 waves of equal and constant amplitudes but temporally varying phases $\varphi_j(t), (j=1,\dots,12)$, (original data: black curve, symbols: reconstruction). (c) Phase functions $\varphi_j(t)$ determined by simulated annealing. Successful reconstruction indicates that a completely linear explanation of rogue wave formation via phase diffusion of a small number of elementary waves is possible. \label{fig:2}}}
\vspace{-1em}
\end{figure}

Nonlinear time series analysis offers a method that allows effective estimation of the number of interfering elementary waves from observations at a single location in the ocean. One suitable method is the Grassberger-Procaccia analysis (GPA) \cite{Grassberger,dimension}, and it delivers an estimate for the dimension $D$ of the parameter space of the system. While the dynamics of simple mechanical oscillators is fully characterized by knowledge of a single phase, the interference of many oscillators at one point on the surface of the ocean, in contrast, is ruled a large number $D=N$ of individual phases, see Figs.~\ref{fig:2}(b,c). Previous studies of ocean wave dynamics indicated values of $D$ reaching from $7$ to $10.5$, with a certain tendency of lower values appearing in calmer waters \cite{Bergamasco,Elgar}. Applying the GPA to wave height measurements recorded on January 1, 1995 on the Draupner platform \cite{Draupner}, we determine a dimension $D$ in the range of $12$ to $13$, see Fig.~\ref{fig:2}(a). The underlying measurement [Fig.~\ref{fig:2}(b)] is the first record of a rogue wave, which ultimately confirmed their existence.  The significant wave height $H_{\rm S}$ in this event was 12\,m, and the rogue wave exhibited a wave height of 25.6\,m trough to crest. For comparison, we also ran the GPA on other data sets \cite{PRL} recorded during the Draupner storm, which provided a nearly identical result for $D$.
It should be noted that significant wave heights were considerably smaller ($H_{\rm S}<4$\,m) in previous studies \cite{Bergamasco,Elgar}. The clearly deterministic signature of ocean wave records \cite{PRL} suggests that the dynamics of the water surface arises due to interference of a relatively small number of elementary waves. Moreover, this number is certainly not a constant and may also depend on factors other than $H_{\rm S}$.

Linear interference of waves with random phase is a well-known assumption for modeling ocean waves \cite{LonguetHiggins}. Similar statistics are relevant for the amplitude distribution of multiply scattered radar signals \cite{IEEE1}. Recently, the emergence of rogue wave statistics due to purely linear interference of optical waves was observed in two different experimental situations \cite{Arecchi,Fratalocchi}, and similar linear rogue waves were seen in microwave experiments \cite{linearmicrowave}. The experiments by Liu {\em et al.} beautifully confirmed the Rayleigh distribution suggested for the limit of an infinite number of interfering waves \cite{Fratalocchi}. In all these examples, the fundamental physics of the random interference phenomenon appears nearly identical, yet with a differing number $N$ of interfering waves. Figure \ref{fig:1}(a) shows simulations of randomly interfering waves on a two-dimensional surface in comparison to the situation of local perfect constructive interference of the waves, see Figs.~1(b-d) and movie sequence in the SI. Depending on the rate of the modeled phase diffusion process \cite{Latifah}, either the characteristic preceding and trailing troughs of the Draupner event emerge [Fig.~1(c)], or a pattern appears that is reminiscent of the famous ``three sisters'' rogue wave \cite{threesisters} [Fig.~1(d)]. Moreover, model parameters can also be adjusted to produce an isolated rogue hole \cite{roguehole} [Fig.~1(b)], which appears to be the most mysterious rogue wave variant reported to date.

\begin{figure}[tb]
\centering
\includegraphics[width=8.5cm]{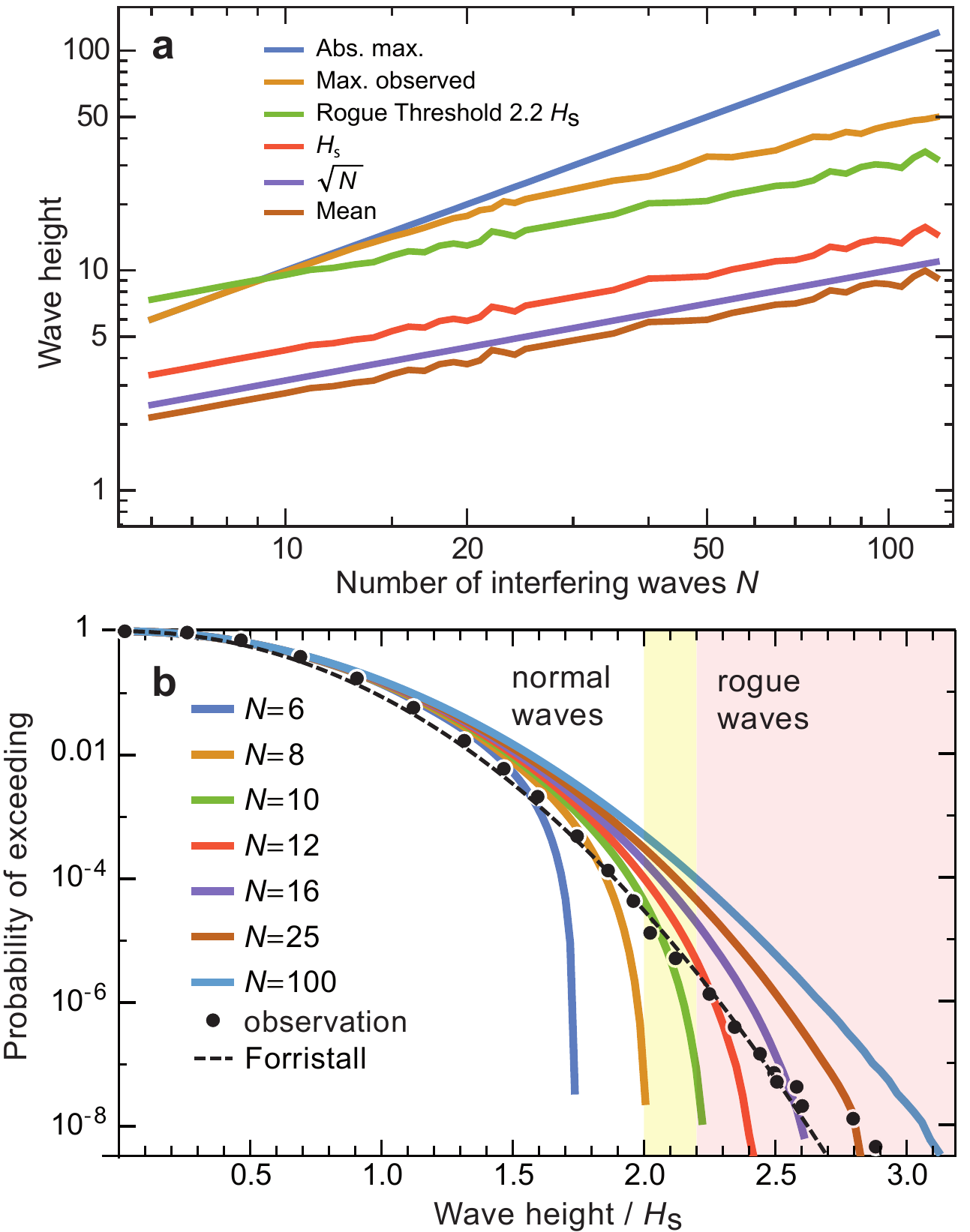}
\vspace{-1em}
\footnotesize{\caption{Simulation of statistical distributions resulting from the interference of $N$ waves with unity amplitude and random phases. For each value of $N$, $10^9$ resulting wave heights have been computed, i.e., about 4 times more than available in Ref. \cite{Christou}. (a) Resulting average wave height, significant wave height $H_{\rm S}$ as well as maximum observed wave height as a function of $N$. For identical phases, a maximum possible wave height results at $N$ times the height of the individual input waves. Using the threshold condition of $2.0$ or $2.2 H_{\rm S}$, no rogue waves are possible below $N=8$ or $10$, respectively. (b) Computed probabilities of exceeding a given wave height in units of $H_{\rm S}$. Cases from $N=6$ to $100$ are shown as lines. Observations from Christou and Ewans are shown as dots \cite{Christou}. The Foristall distribution is shown as a dashed line \cite{Forristall}. The fact that no single simulation fits to the observations is indicative of a strong variation of $N$ due to changing weather conditions during the observations. \label{fig:3}}}
\vspace{-1em}
\end{figure}

Figure \ref{fig:3} shows results of more detailed numerical simulations of the random interference of a finite number $N$ of waves, see Methods section for details.
The resulting mean height amounts to $\sqrt{\pi N}/2$, see Fig.~\ref{fig:3}(a).
The resulting probability densities $p_N(h)$ are skewed, and a heavy tail starts to form with increasing $N$.
We adopt the fairly strict definition that requires rogue waves to exceed the significant wave height $H_{\rm S}$ by a factor 2.2 and find that this condition can only be met for $N>10$,
see Fig.~\ref{fig:3}(a). Using $N = D = 12$ as suggested by the GPA, the probability for rogue wave appearance amounts to
$\mathbb{P}(h>2.2 H_s) \approx 2 \times 10^{-6}$, which agrees favorably with an exhaustive analysis of $10^5$ days of wave height observations \cite{Christou}, see Fig.~\ref{fig:3}(b).
For example, assuming a wave period of 15\,s, these estimates translate into the appearance of a single rogue event within hundreds of days of permanent Draupner storm conditions.
Going to significantly higher values of $N\approx 20$, the probability for rogue wave dramatically increases, with resulting probability density functions slowly converging towards a Rayleigh function.
Under these severe storm conditions, probabilities for the emergence of rogue waves may readily be ten times higher than during the Draupner event ($N=12$). Comparing the simulated probabilities with the analysis of Christou and Ewans \cite{Christou}, nevertheless, it appears striking that the observed trend of wave heights does not follow the simulated probability for any fixed number $N$.
This finding reflects the fact that this extensive wave record was not recorded under constant weather conditions and therefore corresponds to a mixture of probability distributions with various values of $N$.
In fact, the comparison suggests that the ocean dynamics is probably well characterized by $N=6$ throughout most of the time. However, occasional storms seem to even
exceed $N=25$.
A second explanation for the largest observed rogue events $h>2.5 H_{\rm S}$ in the Christou record may be nonlinear steepening or compression effects, which further amplify the heavy tail, similar as in the Tayfun correction \cite{Tayfun} to linear statistics. Such mechanisms have been previously discussed as super rogue waves \cite{Chaboub}.

The variable character of $N$ appears key to understanding the rogue wave phenomenon. As the waves are generated by winds across the ocean surface, one expects to see an increase of the average number of interfering waves, e.g., when wind directions change over a wide range or when turbulence sets in. A strong variability of wind directions is expected to generate waves within a wider angular range and may therefore give rise to larger values of $N$. Moreover, an increasing build-up of large waves on the ocean surface will also serve to increase the turbulent nature of the wind blowing across this undulated surface. The hypothesis of a connection between strong winds and frequent rogue waves is also well confirmed by the observations of Christou and Ewans \cite{Christou} who found hundreds of rogue waves at strong winds, but could not find a single one at below-average wind conditions. Additionally, they also observed an increase of rogue events for the case of a wide angular spread of the swell. Let us further remark that the number of interfering waves may also be influenced by particular coastal shapes, subsea topography, or ocean currents, which may have a focusing or cumulating effect on otherwise spatially separated waves \cite{Current,Focusing}. In order to identify such possibly dangerous regions, we suggest using the correlation dimension $D$ for characterization of the sea state. As indicated by the example analysis of the Draupner data, relatively short time series of 20 minutes length already suffice for an estimate of $D$, and required computational times are on the order of a few minutes. Values above the threshold of $D \approx 10$ indicate the possibility of rogue waves, with a rapidly increasing probability for higher $D$. At open waters far away from any coastline, it may additionally prove useful to carefully monitor the wind situation, in particular the variability of the wind directions or the angular spread of the swell. A further possibility for identifying rogue wave situations may be given by measuring the transverse extent of the
wave crests, which is expected to shrink with increasing angular spread of interfering waves.

In conclusion, our analysis indicates that rogue waves in weakly nonlinear systems may simply appear due to linear interference of a finite number of plane waves larger than ten. No nonlinearity in the oceanic system is therefore required to explain the formation of a heavy tail, yet nonlinearities will certainly give rise to modification of the probability density functions, and they may also explain the extremely steep waveforms reported for oceanic rogue waves. Our findings indicate particular wind conditions as the real origin of rogue waves in the sea, and nonlinearities in the atmospheric system certainly play a decisive role in generating strongly varying wind directions. Quite surprisingly, moderate wind conditions may completely thwart the formation of rogue waves whereas the fully developed turbulence in gale-force storms may invoke these ``monsters of the deep'' at much higher probabilities than previously thought possible. The system size estimate $D$ from the GPA analysis may play a pivotal role
in forecasting rogue wave situations, and it may further help to identify particularly dangerous areas in the ocean. Finally, given the critical dependence on wind conditions, even very minor climate changes \cite{climatechange} strongly affect the probability of rogue waves. A $10\%$ increase of wind speeds in an average storm may readily double the appearance of rogue waves. As the bottom line, the answer to the ocean rogue wave mystery does not seem to lie in the depth of the ocean as is commonly believed; instead, this answer appears to be literally blowing in the wind.


\section*{Methods}
\subsection*{Probability Density Functions}
Various probability distributions been discussed for ocean waves. In the case of an infinite number of linearly interfering waves, a {\em Rayleigh} density function
\begin{equation}
p_{\rm Rayleigh}(h)= \frac{h}{\sigma^2} \exp\left( - \frac{h^2}{2 \sigma^2} \right)
\end{equation}
of wave heights $h$ results \cite{LonguetHiggins}. $\sigma$ is a scale parameter, which parameterizes both the average height
and the standard deviation  of the waves.
Taking nonlinearities of the ocean wave system into account, {\em M.~A.~Tayfun} derived the following correction \cite{Tayfun} to the Rayleigh distribution
\begin{eqnarray}
p_{\rm Tayfun}(h) & = & 2 \beta \left( 1- \frac{1}{\sqrt{1+2 h / \beta}}\right) \nonumber \\
& \times & \exp\left[ - \beta^2 \left( \sqrt{1+2 h / \beta} -1 \right)^2 \right],
\end{eqnarray}
where
\begin{equation}
\beta=\frac{1}{\sqrt{2} \gamma \sigma}
\end{equation}
and $\gamma$ is a nonlinearity parameter, which is typically chosen such that the product $\gamma \sigma$ is in the range from 0.1 to 0.3, further increasing the heavy tail in the distribution.
While both of these distributions have been reported to overestimate the heavy tail, best agreement with observed data was reported for
\begin{equation}
p_{\rm Forristall}(h) = \frac{h}{\sigma^2} \exp\left( - \frac{h^{2.126}}{2.105 \, \sigma^2} \right),
\end{equation}
as was empirically found by {\em G.~Z.~Forristall} \cite{Forristall}. The excellent agreement with the total of 30\,years of wave data was independently confirmed by Christou and Ewans \cite{Christou}.
Using any of the above definitions, the probability $\mathbb{P}(H>h)$ for the wave height $H$ to exceed a given value $h$ is computed as
\begin{equation}
\mathbb{P}(H>h)=\int^\infty_{h} \limits p(s) {\rm d}s \;\mathlarger{\mathlarger{\mathlarger{\mathlarger{\mathlarger{/}}}}}\; \int^\infty_0 \limits p(s) {\rm d}s .
\end{equation}

\subsection*{Grassberger-Procaccia Dimensional Analysis}
Neglecting dynamical changes in the amplitudes, the dynamics of the ocean system at one fixed location is ruled by a number of $N$ individual phases, which is identical to the dimension $D$ of the parameter space, i.e., $D=N$. The number $D$ can be estimated by the Grassberger-Procaccia algorithm\cite{Grassberger} (GPA),  which is a standard method of nonlinear time series analysis. The GPA analyzes a time series of a physical quantity $x$. This time series has a length $N$ and is discretely sampled with a constant sampling rate. We denote the time series as $\vec{x} = \{x_1, x_2, x_3, \ldots, x_N\}$. From this time series $\vec{x}$, sub-series $\vec{y}_{i} = \{x_i, x_{i+1}, x_{i+2}, \ldots, x_{i+m}\}$ of length $m$ are selected. The length $m$ is called the \textit{embedding dimension}. Each sub-series $\vec{y}_{i}$ is compared to all sub-series $\vec{y}_{j}$ with $j>i$ by calculating Euclidian distances
\begin{eqnarray}
	r_{ijm} = \parallel \vec{y}_{i} - \vec{y}_{j} \parallel = \sqrt{\sum^{i+m}_{k=i} |x_k - x_{k+j-i}|^2}.
\end{eqnarray}
For $m \ll N$ the Euclidian distances $r_{ijm}$ are accumulated in the \textit{correlation sum} \begin{eqnarray}
	C_m(r) = \frac{1}{N^2} \sum_{j>i} \theta(r - r_{ijm}),
\end{eqnarray}
where $\theta(r)$ is the Heaviside step function. In order to avoid false interpretation due to detection noise and limited sample length, $C_m(r)$ is only analyzed at the interval $[0.01~r_{\rm{max}}, 0.1~r_{\rm{max}}]$, where $r_{\rm{max}}$ is the largest Euclidian distance found in the data series. The correlation sum $C_m(r)$ increases monotonically according to
\begin{equation}
C_m(r) \propto r^\nu,
\end{equation}
with the exponent $\nu$.
For small values of $m$, $\nu$ increases with embedding dimension $m$ until it reaches a saturation value $d_{\rm{sat}}$. For a range of values of $m$, a plateau with nearly constant $d_m=d_{\rm{sat}}$ is typically observed (cf.~Fig.~\ref{fig:2}), provided a sufficient number of data points in the analysis. Strictly speaking, the correlation dimension $d_{\rm{sat}}$ delivers a lower bound for the (Hausdorff) dimension $D$ of the parameter space of the system\cite{Grassberger}. In many cases, $d_{\rm{sat}}$ directly delivers a good estimate for the dimension $D$. In case of a number of $N$ interfering mechanical oscillators of constant amplitude, the parameter space of the system is composed of $N$ independent phases $\varphi_N$. In this situation, we can set $D=N \gtrsim d_{\rm{sat}}$.

\subsection*{Superposition of Waves as a Two-Dimensional Random Walk Problem}
The problem of the resulting amplitude of $N$ interfering waves is mathematically equivalent to computing the probability of the length $h$ of two-dimensional random walks\cite{IEEE1,IEEE2,LonguetHiggins} involving $N$ steps
\begin{equation}
h = \left| \sum_{k=1}^N h_k \exp( i \varphi_k ) \right|,
\end{equation}
where $\varphi_k \in [0,2 \pi [$ are univariate uncorrelated random phases and $h_k$ are the heights of the individual waves. Assuming $h_k \equiv h_0$, the absolute maximum possible resulting wave height is $N h_0$, and for sufficiently large $N$ the mean height is $\approx \sqrt{\pi N} h_0 / 2$, cf.~Fig.~\ref{fig:3}(a). The threshold for rogue waves scales as $\approx 3.1 h_0 \sqrt{N}$. If all $\varphi_k$ are identical, the maximum possible amplitude of $h_{\rm max}=N h_0$ occurs. Comparing the rogue threshold with the maximum possible wave height, it becomes immediately clear that a minimum of 10 linearly interfering waves is required to exceed the rogue threshold. The resulting probability densities $p_N(h)$ are skewed, and a heavy tail starts to form with increasing $N$. Computing $h$ for a large number of randomly chosen $\varphi_k$ enables the computation of probability density functions for a given $N$. For the examples shown in Fig.~\ref{fig:3}, a total of $10^9$ random sets has been employed for any of the shown values of $N$.

\subsection*{Stochastic Model of Rogue Wave Emergence}
In numerical simulations, we assumed linear interference of $N$ elementary waves\cite{LonguetHiggins} of unity amplitude according to
\begin{equation}
h(t,\vec{r})= \sum_{j=1}^N \cos\left( \vec{k}_j \vec{r} + \omega t + \varphi_j(x,t) \right).
\end{equation}
Here a constant angular frequency of the waves $\omega$ has been assumed. $t$ is the time, and $\vec{r}=(x,y)$ is the spatial coordinate.
The wave vectors are defined as $\vec{k}_j=(\cos(\theta_j),\sin(\theta_j))$, with a Gaussian distribution of random angles $\theta_j$ of vanishing mean. The standard deviation $\Delta\theta$ of the angular distribution of waves determines the lateral extent of a rogue wave along the $y$ direction. Using a $2 \pi$ univariate angular distribution instead, one can simulate the emergence of circular symmetric rogue waves as observed by Arecchi {\em et al}\cite{Arecchi}.
A one-dimensional diffusion process of the phase functions $\varphi_j(x,t)$ along the propagation direction\cite{Latifah} is computed at equidistant positions $x_i=i \Delta_x$ according to the recursive law
\begin{equation}
\varphi_j(x_{i \pm 1},t)=\varphi_j(x_{i},t)+D_\varphi \tilde\varphi_{ij}. \label{recursion}
\end{equation}
with a diffusion coefficient $D_\varphi$ and a set of independent Gaussian random phases $\tilde\varphi_{ij}$ with mean zero. This model assumption induces a random phase walk, which nevertheless leaves the average frequency $\omega$ unaffected. The characteristic length or time over which dephasing appears is related to the prediction length or linear correlation time\cite{PRL}. Rogue waves appear by setting $\varphi_j = 0$ at $t=0$ and $x=0$; rogue holes can be generated by forcing all $\varphi_j = \pi$.

The above model can also be inverted to find a set of phase functions $\varphi_j(t)$ that give rise to a measured wave record $h(t)$, cf.~Figs.~\ref{fig:2}(b,c). Starting at the time $t=0$ of the rogue event, we used a simple simulated annealing variant to adjust the individual phases $\varphi_j(t)$ in a round-robin fashion to the end of obtaining best agreement with the measured record of the Draupner event. This procedure is repeated for each time step in the measured time series. As this phase retrieval is an ill-posed problem, rapid temporal oscillations of the phase functions may locally appear. These oscillations can be mostly suppressed by a suitable penalty term, with the noted exception of the deep trough areas of the Draupner wave. This may be an indication for unaccounted nonlinear shaping in the immediate vicinity of the rogue wave. It should also be noted that phase retrieval is certainly highly ambiguous in this situation, but nevertheless indicates that rogue wave formation can be explained due to phase diffusion in the interference of a finite number of elementary wave with constant and equal amplitude. The phase dynamics near $t=0$ shows essentially a linear dephasing behavior with time, mostly being caused by differing frequencies of the interfering waves. For the Draupner event, the phase diffusion rate between the individual waves amounts to $4 \pi$ per minute.

Temporal phase diffusion can be simulated in a completely analogous way ($t_i=i \Delta_t$) by
\begin{equation}
\varphi_j(x,t_{i \pm 1})=\varphi_j(x,t_i)+D_t \tilde\phi_{ij} \label{recursion2}
\end{equation}
with Gaussian random phases $\tilde\phi_{ij}$ and an independent diffusion constant $D_t$. In Fig.~\ref{fig:1}, $N$ has been set to 20, which exceeds the estimate from the dimensional analysis ($D=12$ to 13). For simulation of a Draupner-like event, the diffusion constant $D_\varphi$ was chosen such that dephasing appears on the scale of one oscillation period according to the above consideration. For the ``three sisters,'' smaller values of $D_\varphi$ need to be used, giving rise to dephasing after more than one oscillation cycle. Provided a sufficient number of rogue wave data sets, the exact value of the diffusion constant can be estimated from a linear correlation analysis of the wave record. The angular spread of the $\theta_j$ has been adjusted to 50\,mrad to allow a certain vertical extension of the rogue waves. Narrowing down the spread, wider and wider walls of water appear. Parameter $D_t$, finally, determines how long a rogue wave will survive while propagating on the ocean surface. At the current stage, we are not aware of any measurements that would allow estimating this parameter.

Accounting for the nonlinearity of the ocean system, one can include a nonlinear correction in the simulation, which effectively reduces the depth of the troughs and increases the crests. When doing so, a substantially larger number $N$ of interfering waves is required to form a rogue hole whereas the likelihood of rogue holes and waves is equal in a completely linear model without correction.

Our model relates the emergence of ocean rogue waves to only three parameters, namely $N$, $D_\varphi$, and $D_t$. The effective number $N$ of interfering waves can be estimated via the GPA analysis, the phase diffusion coefficient $D_\varphi$ can be related to the characteristic number of crests and troughs in a rogue event or to a linear correlation analysis of the time series. The phase diffusion rate in time is currently unknown, but this process is expected to be slower than the primary phase diffusion, as rogue waves would otherwise only remain stable for a few seconds. Among the three mentioned parameters, $N$ is the most important one, since its knowledge allows predicting probabilities of wave heights by isolating the dynamics at one fixed point on the ocean surface.

\section*{Acknowledgments}
 We gratefully acknowledge financial support by Deutsche Forschungsgemeinschaft (DFG STE 762/9) and Nieders.~Vorab ZN3061. We thank Statoil ASA for the permission to use the data recorded on the Draupner oil platform in January 1995. We acknowledge fruitful discussions with Nail Akhmediev, Amin Chabchoub, Bernd Hofmann, and Miro Erkintalo.



\end{document}